\documentclass[12pt,a4paper]{article}
\begin{document}

\title{Can many-valued logic help to comprehend quantum phenomena?}
\author{Jaros{\l}aw Pykacz\thanks{Institute of Mathematics, University of Gda\'nsk, Poland; e-mail: pykacz@mat.ug.edu.pl}}
\date{}
\maketitle
\thispagestyle{empty}

\begin{abstract}

Following {\L}ukasiewicz, we argue that future non-certain events should be described with the use of many-valued, not 2-valued logic. The Greenberger - Horne - Zeilinger `paradox' is shown to be an artifact caused by unjustified use of 2-valued logic while considering results of future non-certain events. Description of properties of quantum objects before they are measured should be performed with the use of propositional functions that form a particular model of $\infty $-valued {\L}ukasiewicz logic. This model is distinguished by specific operations of negation, conjunction, and disjunction that are used in it.\\


\end{abstract}
\section{Introduction}

In the history of physics there were already several moments when stubborn sticking to old and up to that date efficient ideas or models could have stopped the development of our understanding of the Reality. Such moments were often marked by emergence of `paradoxes' yielded by old theories. The relief was usually obtained through radical change of some `fundamental' ideas that had been, up to that date, considered as indisputable.

Let us mention some instances of such revolutionary changes that allowed to make progress that otherwise could not be possible.

\begin{enumerate}
    \item The Copernicus revolution that consisted in radical change of the adopted model of the Universe.
    \item Emergence of Special Relativity could not be possible without abandoning linear addition of velocities and assuming that the velocity of light is the same in all inertial frames.
    \item The idea that numerous physical quantities considered previously as continuous--valued can assume only discrete values was the cornerstone of quantum mechanics.
    \item Emergence of General Relativity was connected with abandoning the idea that geometry of physical space is a flat Euclidean one.
\end{enumerate}

The aim of the present paper is to argue that abandoning two-valued logic in favor of many-valued one while considering properties of quantum objects \emph{prior to their measurements}, although as `revolutionary' as the changes of paradigms listed above, could be indispensable in order to to free quantum mechanics from some of its recently widely discussed paradoxes. We strengthen our arguments by showing the natural way in which many-valued logic enters mathematical description of properties of quantum objects.

\section{Bell, Kochen - Specker, GHZ, and all that...}

The first attempts to explain numerous paradoxes of quantum mechanics on the basis of many-valued logic were made by Polish logician Zygmunt Zawirski in the early Thirties of the XX Century. Zawirski in his papers published in 1931 \cite{Zaw31} and in 1932 \cite{Zaw32} argued that Bohr's complementarity could be comprehended only on the ground of at least 3-valued logic since according to classical 2-valued logic a proposition `\emph{light is a wave and light consists of particles}' is a conjunction of two propositions that cannot be simultaneously true, therefore it is false. However, according to 3-valued {\L }ukasiewicz logic \cite{Luk20} if we ascribe to both these propositions the third truth value: 1/2, interpreted as `possible', their conjunction according to the formula postulated by {\L }ukasiewicz $v(p\wedge q)=min[v(p),v(q)]$ has also truth value 1/2.

The idea that some version of many-valued logic could be useful to explain quantum phenomena was further on pursued by Fritz Zwicky \cite{Zwi33}, Paulette Destouches-F\'{e}vrier \cite{Fev51, Fev37}, Hans Reichenbach \cite{Rei44, Rei48, Rei51, Rei53}, Hilary Putnam \cite{Put57} and Carl Friedrich von Weizs\"{a}cker \cite{Wei58}. Later on this idea was abandoned for some time what Max Jammer observed in his famous book \cite{Jam74}: `After von Weizs\"{a}cker's work no serious attempt seems ever to have been made to elaborate further a many-valued logical approach to quantum mechanics'. One of the aims of the present paper is to argue that today Jammer's words are no longer tenable.

The `paradoxes' that are concerned in this paper all could be named `pre-existence of properties paradoxes'. We choose the Greenberger-Horne-Zeilinger `paradox' \cite{GHZ} (see also \cite{Mer90}) as a typical representative of them.

\subsection{The GHZ `paradox': classical derivation}

Essentially, the derivation of the GHZ `paradox' boils down to the following argumentation: Let us consider three spin-1/2 particles in the entangled state $|\psi \rangle = \frac{1}{\sqrt{2}}|111 \rangle + |000 \rangle$ and let spin measurement either in the $X$ direction or in the $Y$ direction is performed on each particle. Let us ascribe to the result of the measurement of spin of the $i$-th ($i = 1,2,3$) particle in the $X$ or $Y$ direction the number $+1$ or $-1$, depending whether the result of the measurement is `up' or `down', respectively. Quantum-mechanical calculations show that products of these numbers have to fulfill the following equations:
\begin{eqnarray}
X_1Y_2Y_3 &=& +1\\
Y_1X_2Y_3 &=& +1\\
Y_1Y_2X_3 &=& +1\\
X_1X_2X_3 &=& -1.
\end{eqnarray}
Let us now assume that each of the quantum particles, before the measurement is done, actually possesses all properties that are later on revealed by the measurement, i.e., that we can ascribe to $X_i$, $Y_i$ the numbers $+1$ and $-1$ in the unique way. This assumption is in a contradiction with the above-stated results of quantum-mechanical calculations since all four equations (1) - (4) cannot hold simultaneously: The product of all left-hand sides is $(X_1Y_2Y_3)(Y_1X_2Y_3)(Y_1Y_2X_3)(X_1X_2X_3) = X_1^2X_2^2X_3^2Y_1^2Y_2^2Y_3^2 = +1$ while the product of all right-hand sides is $1\cdot 1\cdot 1\cdot(-1) = -1$. Therefore, since all experiments performed up-to-day confirm predictions of quantum mechanics, the claim that quantum objects possess all properties prior to their measurements seems to be untenable. This position was eloquently expressed by Peres in the title of his paper \cite{Per78}: `\emph{Unperformed experiments have no results}'. However, the reasoning presented above crucially depends on the assumption that all the statements of the form: `\emph{spin of the object i in the direction $X_i$ ($Y_i$) is up (down)}' are either entirely true or entirely false, i.e., that they belong to the domain of 2-valued logic. If in this description 2-valued logic is replaced by many-valued logic, the GHZ `paradox' could not be derived. In order to make usage of 2-valued logic in the derivation of the GHZ `paradox' clearly visible, we present in the next subsection the equivalent derivation of the GHZ `paradox' that highlights the role of the underlying logic.

\subsection{The GHZ `paradox': derivation based explicitely on 2-valued logic}

Let us consider quantum-mechanical equations (1) - (4), replace numerical values $+1$ and $-1$ by truth values of sentences: $\bf{X_i}$ ($\bf{Y_i}$) =
 `\emph{spin of the object i in the direction $X $ ($Y $) is up}', and algebraic product of numbers by `exclusive OR' operation (XOR) that will be denoted $\oplus $ in the sequel. Since XOR operation in 2-valued classical logic is associative, no brackets are needed in the expressions of the form $\bf{X_1} \oplus \bf{Y_2} \oplus \bf{Y_3}$, etc. Moreover, truth values of compound sentences of the form $\bf{A_1} \oplus \bf{A_2} \oplus ... \oplus \bf{A_n}$ depend soleley on the parity of their true elementary constituents: The truth value of such a compound sentence is $1$ iff the number of its true elementary constituents is odd, $0$ otherwise. Taking this into account, it is straightforward to check that quantum-mechanical equalities (1) - (4) are equivalent to the following expressions valid in classical 2-valued logic:
\begin{eqnarray}
\bf{X_1} \oplus \bf{Y_2} \oplus \bf{Y_3} &\equiv& V\\
\bf{Y_1} \oplus \bf{X_2} \oplus \bf{Y_3} &\equiv& V\\
\bf{Y_1} \oplus \bf{Y_2} \oplus \bf{X_3} &\equiv& V\\
\bf{X_1} \oplus \bf{X_2} \oplus \bf{X_3} &\equiv& F,
\end{eqnarray}
where $V$ and $F$ denote, respectively, the true and the false propositions, and $\equiv $ means `\emph{has the same truth value as}'.

Now, if we form exclusive disjunction of all left-hand sides and all right-hand sides of these expressions, we obtain contradiction since on the left-hand side of the formula obtained in such a way each term appears twice, so its truth value is 0, while $v(V \oplus V \oplus V \oplus F) = 1$.

Let us stress, that the contradiction could not be obtained without assuming that each statement of the form $\bf{X_i}$ ($\bf{Y_i}$) =
 `\emph{spin of the object i in the direction $X $ ($Y $) is up}' has definite truth value that is either $0$ or $1$ before suitable measurement is done, and that laws of classical 2-valued logic hold.

It is obvious that also in the case of other `paradoxes' which follow from the assumption that quantum objects possess all properties prior to their measurements this `possessment' is tacitly assumed to be `absolute', i.e., that any sentence of the form `\emph{an object i has property P}' is assumed to be either entirely true or entirely false. In some cases, e.g., derivation of Kochen - Specker theorem, this assumption is made explicitely.

\section{Many-valued logic is a proper tool for description of not-yet-measured properties of quantum objects}

Jan {\L }ukasiewicz, the founding father of modern many-valued logic argued in his address which he delivered as a Rector of Warsaw University at the Innaguration of the academic year 1922/1923 \cite{Luk22} that `\emph{All sentences about future facts that are not yet decided belong to this} [many-valued] \emph{category. Such sentences are neither true at the present moment ... nor are they false...}'. This attitude is clearly applicable to sentences concerning the results of future experiments, which are not decided at present. Let us note the full analogy between often by {\L }ukasiewicz quoted Aristotle's statement `\emph{There will be a sea battle tomorrow}' and quantum mechanical predictions of the form `\emph{A photon will pass through a filter}'. In both cases the position of classical 2-valued logic is such that since the occurence or non-occurence of these events is not certain, these statements cannot be endowed with any truth value, i.e., they do not belong to the domain of 2-valued logic. However, it was shown in the previous Section that this forbidden procedure was applied in derivation of the GHZ `paradox'. Of course the reason for which it was treated as being allowed was the assumption that the results of future measurements are pre-determined, as it happens in classical mechanics where the usage of classical 2-valued logic is fully justified. However, let us note that classical mechanics describes Imaginary World, not Real World that we live in and which, to the best of our contemporary knowledge, is governed by laws of quantum mechanics. Therefore, changing classical 2-valued logic to many-valued logic while discussing future non-certain events seems to be inevitable -- otherwise we could not reasonably talk about them at all.

Let us note that in fact we use many-valued logic in such situations in everyday life. When one says `\emph{There will be a sea battle tomorrow}', one usually intuitively ascribes some degree of probability (possibility?, likelihood?) to this future event. This is exactly the way in which {\L }ukasiewicz interpreted truth values of statements about future events that are different form $0$ and $1$ in his numerous papers \cite{Luk70}\footnote{The idea that probabilities should be interpreted as truth values of many-valued logic was for the first time expressed by {\L }ukasiewicz already in 1913 in \cite{Luk13}.}.

While considering future events in the quantum realm we are in much better situation because these truth values are known with absolute precission: they are simply probabilities that these events will happen obtained from the theory by exact calculations and the Born rule. Let us note that when an experiment is finished and its results are known with certainty we are back to classical 2-valued logic -- exactly the same happens in macroscopic world (sea battle either happened or it did not happen).

However, in macroscopic world usually the process of converging of many-valued truth values to the extreme values $0$ or $1$ can be seen `dynamically' and can be traced in a more detailed way: when two hostile navies approach each other, truth value of a sentence `\emph{There will be a sea battle}' tends to unity and finally attains it at the moment that sea battle begins. On the contrary, if navies are more and more distant, truth value of the considered sentence tends to $0$. In the quantum realm such detailed description of the process of passing from many-valued to 2-valued logic is not provided.

Having argued so much in favor of many-valued logic as a proper tool for describing results of future non-certain events by quantum theory, we are left with a fundamental question: Which particular model of such logic (and there are infinitely many of them!) should be chosen for this purpose. The problem is a serious one since declaring only that truth values of statements concerning results of future experiments belong to the interval $[0,1]$ instead to a 2-element set $\{0,1\}$ is far too less to specify the model. In particular one would like to know how to calculate truth values of compound sentences when truth values of their constituents are known. In this respect it is worth to mention that although there are only $2^{(2^2)} = 16$ possible binary connectives in 2-valued logic, there are $3^{(3^2)} = 19683$ of them in 3-valued logic, $4^{(4^2)} = 4294967296$ in 4-valued logic and obviously infinity in $\infty $-valued logic. Therefore, the problem of choosing out of them particular ones that are desired generalizations of binary connectives of 2-valued logic, especially of conjunction and disjunction, is a really serious one. In the next section of this paper we shall give some hints for a possible solution of this problem.

\section{Many-valued representation of Birkhoff - von Neumann `quantum logic'}

Since Birkhoff and von Neumann 1936 paper \cite{BvN} it is generally accepted that a family of all experimentally verifiable `yes-no' propositions pertaining to a quantum system is represented by a family of all closed linear subspaces $L(\cal{H})$ of a Hilbert space $\cal{H}$ used to desribe this system. From the algebraic point of view such families are orthomodular lattices, i.e., a non-distributive generalizations of Boolean algebras. Therefore, orthomodular lattices or their slight generalizations: orthomodular partially ordered sets are usually called `quantum logics'. An interested reader is referred to any textbook on quantum logic theory (see, e.g., \cite{BC} or \cite{PP}) for presise mathematical definitions of these notions.

Since according to generally accepted interpretation elements of a quantum logic represent statements about physical systems that, when a suitable experiment is completed, occur to be either true or false, quantum logics are usually seen as 2-valued logics that are, however, non-classical because of lack of distributivity. However, the present author showed in a series of papers \cite{Pyk94, Pyk00, Pyk10, Pyk11} that any physically sound quantum logic in a Birkhoff - von Neumann sense can be isomorphically represented as a family of $\infty $-valued propopsitional functions defined on the set of states of a physical system under study. This family has to fulfill some specific conditions but before we list them we have to introduce some notions.

Let $v(p)$ denote truth value of a proposition $p$. We admit $v(p)$ to be any number from the interval $[0,1]$.

\medskip

\noindent {\bf Definition 1.}
Let $p$, $q$ be propositions. Propositions denoted $\neg p$, $p \sqcap q$, $p \sqcup q$ such that
\begin{eqnarray}
v(\neg p) &=& 1 - v(p)\\
v(p \sqcap q) &=& \max [v(p)+ v(q) - 1,0]\\
v(p \sqcup q) &=& \min [v(p)+ v(q), 1],
\end{eqnarray}
will be called, respectively, \emph{{\L }ukasiewicz negation} of $p$, \emph{{\L }ukasiewicz conjunction} of $p$ and $q$, and \emph{{\L }ukasiewicz disjunction} of $p$ and $q$\footnote{Actually, {\L }ukasiewicz never defined explicitly conjunction (10) and disjunction (11) in his papers. However, we follow terminology that is widely used in contemporary literature.}.

\medskip

Let us note that the triple of operations $(\neg , \sqcap, \sqcup )$ is De Morgan triple, i.e., that for any $p$, $q$
\begin{eqnarray}
\neg (p \sqcap q) &\equiv& \neg p \sqcup \neg q\\
\neg (p \sqcup q) &\equiv& \neg p \sqcap \neg q.
\end{eqnarray}
Moreover, conjunction (10) and disjunction (11) satisfy both the law of excluded middle:
\begin{eqnarray}
v(p \sqcup \neg p) = \min [v(p)+1-v(p), 1] = 1,
\end{eqnarray}
and the law of contradiction:
\begin{eqnarray}
v(p \sqcap \neg p) = \max [v(p)+1-v(p)-1, 0] = 0,
\end{eqnarray}
that are not satisfied by operations originally chosen by {\L }ukasiewicz as many-valued conjunction: $v(p\wedge q) = \min [v(p), v(q)]$ and disjunction: $v(p \vee q ) = \max [v(p), v(q)]$. Therefore, in many-valued logic equipped with the triple of operations $(\neg , \sqcap, \sqcup )$ disjunctions of the form: `\emph{There will be a sea battle tomorrow or there won't be a sea battle tomorrow}' or `\emph{Photon will pass through a filter or it won't pass through a filter}' are always certainly true (i.e., their truth value equals 1) whatever are truth values of propositions that form these disjunctions.

\medskip
\noindent {\bf Definition 2.} Two propositions $p$, $q$ such that $p \sqcap q \equiv F$ will be called \emph{exclusive}.
\medskip

Let us note that since we are in the realm of many-valued logic, exclusiveness of two propositions does not necessarily mean that at least one of them is certainly false. It is straightforward to check that $p \sqcap q \equiv F$ means that $v(p) + v(q) \leq 1$. For example, any proposition and its {\L }ukasiewicz negation are exclusive, but nevertheless they can have non-zero truth values.

We shall deal in the sequel with many-valued propositional functions defined on the set of ststes of a physical system under study, i.e., objects that become many-valued propositions when (a name of) a specific state is inserted into them. Among such funstions there will be only two constant functions (i.e., in fact propositions): The always-false function $F$ and the always-true function $V$. Of course all notions introduced above can be in a straighforward (`pointwise') way extended to propositional functions. For example, the negation of a function $p(\cdot )$ is a function $\neg p(\cdot )$ such that $v(\neg p(|\psi \rangle )) = 1 - v(p(|\psi \rangle ))$ for any state $|\psi \rangle $.

Now we can quote already announced theorem about many-valued representation of Birkhoff - von Neumann quantum logics. This theorem was proved as fuzzy set representation in \cite{Pyk94} and later on it was `translated' to be many-valued representation in \cite{Pyk00}, see also \cite{Pyk10}. Since we are particularly interested in representation of a specific Birkhoff - von Neumann quantum logic: $L(\cal{H})$, we confine our presentation to this particular case.

\medskip

\noindent {\bf Theorem.} Let $L(\cal{H})$ be a family of all closed linear subspaces of a Hilbert space $\cal{H}$ used to desribe a studied quantum system, and let $S^1 (\cal{H})$ be the set of all unit vectors in $\cal{H}$. $L(\cal{H})$ can be isomorphically represented as a family $\cal{L}(\cal{H})$ of $\infty$-valued propositional functions defined on $S^1 (\cal{H})$ that satisfies the following conditions:
\begin{enumerate}
    \item $\cal{L}(\cal{H})$ contains the always-false proposiotional function $F$.
    \item $\cal{L}(\cal{H})$ is closed with respect to {\L }ukasiewicz negation (9), i.e., if $p(\cdot) \in \cal{L}(\cal{H})$, then $\neg p(\cdot ) \in \cal{L}(\cal{H})$.
    \item $\cal{L}(\cal{H})$ is closed with respect to {\L }ukasiewicz disjunction (11) of pairwisely exclusive propositional functions, i.e., if $p(\cdot )_i \in \cal{L}(\cal{H})$ and $p(\cdot )_i \sqcap p(\cdot )_j \equiv F$ for $i \neq j$, then $\sqcup _i p(\cdot )_i \in \cal{L}(\cal{H})$.
    \item The always-false propositional function $F$ is the only propositional function in $\cal{L}(\cal{H})$ that is exclusive with itself, i.e., for any $p(\cdot ) \in \cal{L}(\cal{H})$ if $p(\cdot ) \sqcap p(\cdot ) \equiv F$, then $p(\cdot ) \equiv F$.
\end{enumerate}

\medskip

If we assume, as it is done in traditional quantum logic theory (see, e.g., \cite{BC}, \cite{PP}), that elements of $L(\cal{H})$ faithfully represent properties of a physical system described with the aid of $\cal{H}$, and elements of $S^1 (\cal{H})$ faithfully represent its pure states, one is led to the following interpretation of the propositional functions mentioned in this Theorem:

Let $P$ be a property of a quantum object, e.g., a photon's property of being able to pass through a filter. A sentence `\emph{A photon will pass through a filter}' is obviously a propositional function, since it attains a definite truth value only when the phrase `\emph{in a state represented by  $|\psi \rangle $}' is inserted into it. Let us denote it $p(\cdot )$, while $p(|\psi \rangle )$ denotes the $\infty $-valued proposition `\emph{A photon in a state represented by $|\psi \rangle $ will pass through a filter}'.

Let $\hat{P}$ be an operator of orthogonal projection onto a closed linear subspace of $\cal{H}$ that, according to the traditional approach to quantum logics, represents the property $P$. Then the truth value of the proposition $p(|\psi \rangle )$, according to the discussion presented in previous sections of this paper, equals to the probability that a result of an experiment designed to check whether a photon in a state $|\psi \rangle $ possesses the mentioned property, will be positive. Since according to the orthodox quantum theory this probability equals $ \langle \psi |\hat {P}|\psi \rangle $, we are bound to accept it as a truth value of the proposition $p(|\psi \rangle )$:
\begin{eqnarray}
v(p(|\psi \rangle )) =  \langle \psi |\hat {P}|\psi \rangle .
\end{eqnarray}
Let us note that because of the close link between many-valued logic and fuzzy set theory, that is as close as the link between classical 2-valued logic and traditional theory of crisp sets, one can assume that the number (16) represents degree to which a quantum object possesses a property $P$ before an experiment designed to check this property is completed. Therefore, we can paraphrase the title of Peres' paper \cite{Per78} and say that: \emph{Unperformed experiments have all their possible results, each of them to the degree obtained from suitable quantum-mechanical calculations}. In particular, a quantum object in a state $|\psi \rangle $ both possesses to the degree $v(p(|\psi \rangle ))$ a property $P$ and does not possess it to the degree $1 - v(p(|\psi \rangle ))$. For example, each of quantum objects considered in the GHZ `paradox' before its spin is measured has a property of \emph{having spin up in the direction X (Y)} to the degree 1/2, and at the same time it has not this property to the same degree. Of course if such a point of view is accepted, the derivation of this `paradox' that is based on 2-valued logic and the idea that each property can be only either entirely possessed or entirely not possessed, cannot be performed.

Let us finish with two macroscopic examples of properties that are neither entirely possessed nor entirely not possessed by an object.

In the first example let us consider Aristotle's two hostile navies $N$ and $M$ that will be involved in a sea battle tomorrow. If we evaluate, taking into account the number of ships, quality of weapon, previous battles, weather conditions, etc., that the probability (possibility? likelihood?) that the navy $N$ will win tomorrow is, say, $0.8$, then we can declare \emph{today} that $N$ has today a property of `\emph{being victorious}' to the degree $0.8$, and of course at the same time it has a property of `\emph{not being victorious}' to the degree $0.2$.

The second example is based on the archetypal situation often considered in fuzzy set theory. Let us agree that a man with $N$ hair definitely belongs to the \emph{set of bald men} iff $N \leq 100$, definitely does not belong to it iff $N \geq 1000$, and that if the number of his hair $N \in (100, 1000)$, then the degree of his membership to the \emph{set of bald men} is a linear function of N, i.e., that in this case
\begin{eqnarray}
\mu (N) = \frac{1000 - N}{900}.
\end{eqnarray}
This degree of membership can of course be treated as a degree to which a man with $N$ hair has a property of \emph{being bald}, or as a truth value of a proposition `\emph{This man is bald}'. Let us note that although in this example we are not dealing with sentences concerning future non-certain experiments, the considered situation can be seen in this way: Let us agree that future `yes-no' experiments with non-certain results will consist in asking randomly chosen people whether a man with $N$ hair is bald or not and forcing them to give only one of allowed answers. In such a situation we are back to the familiar situation of properties that are judged to be either entirely possessed or entirely not possessed only when a suitable experiment is performed.

\section{Summary}

We argued in a paper, following {\L }ukasiewicz, that future non-certain events should be described with the use of many-valued, not 2-valued logic. This concerns both macroscopic and microscopic phenomena. Numerous `pre-existence of properties paradoxes', like the Greenberger - Horne - Zeilinger `paradox' are artifacts caused by unjustified use of 2-valued logic while considering results of future non-certain experiments. Our arguments were strengthen by showing that usual description of properties of quantum objects provided by Birkhoff - von Neumann concept of a quantum logic can be equivalently performed with the use of propositional functions that form a particular model of $\infty $-valued {\L }ukasiewicz logic. The specific operations of negation, conjunction, and disjunction that are used to build this model distinguish it from the infinite family of possible models of many-valued logics.

\medskip

\noindent
\textbf{Acknowledgments.} Financial support of Polish National Center for Science (NCN) under the grant 2011/03/B/HS1/04573 is gratefully acknowledged.

\end{document}